\documentclass[%
aps,%
12pt,%
final,%
notitlepage,%
oneside,%
onecolumn,%
nobibnotes,%
nofootinbib,%
superscriptaddress,%
noshowpacs,%
showkeys,%
centertags%
]{revtex4}
\usepackage{textcomp}
\usepackage{caption2}
\usepackage{epsfig}
\usepackage{graphics}

\pdfoutput=1

\begin{document}

\title{Non-uniform horizons in Gauge/Gravity Duality}

\author{T.M. Moskalets}
\email{tatyana.moskalets@gmail.com}
\affiliation{
School of Physics and Technology, V.N. Karazin Kharkiv National University, 31 Kurchatova Av., Kharkiv, 61108, Ukraine\\
}
\author{A.J. Nurmagambetov$^{1,}$}
\email{ajn@kipt.kharkov.ua}
\affiliation{A.I. Akhiezer Institute for Theoretical Physics of NSC KIPT,\\
1 Akademicheskaya St., Kharkiv, 61108 Ukraine
}

\begin{abstract}
In this communication, based on our paper \cite{Moskalets:2014hoa}, we discuss a way of enhancing Gauge/Gravity Duality and response of a dual strongly coupled medium on placing the inhomogeneity on the gravity side.

\end{abstract}
\keywords{AdS/CFT correspondence, hydrodynamic limit, transport coefficients, strongly coupled media}

\maketitle

\vspace{-1cm}
\section{Introduction}
Recent progress in description of strongly coupled gauge theories in terms of their gravitational duals \cite{Aharony:1999ti} generated a new wave of interest in ``Applied String Theory''. Further steps in this direction revealed non-trivial relations between String Theory and Condensed Matter Physics, in particular, in effective descriptions of a Quark-Gluon Plasma (QGP) \cite{Policastro:2002se}
and high-temperature superconductivity \cite{Hartnoll:2008vx}. However, existence of universal bounds on transport coefficients in strongly coupled dual media \cite{Kovtun:2004de},\cite{Kovtun:2008kx} sharply indicates a rigidity of the hydrodynamical limit of AdS/CFT correspondence in a wide range of dual gauge models, including non-conformal ones. This 
fact greatly narrows the application of Gauge/Gravity Duality to real physical systems, so the question is: is there a way to enhance the Gauge/Gravity Duality to make it more flexible?

Technically, description of a strongly coupled gauge model at non-zero temperature and with non-zero finite density of charges/chemical potentials is based on the appropriate choice of dual gravitational theory.
Physical properties of the effective medium, encoded in solutions to its dynamical equations on the gauge theory side, are intimately related to properties of solutions to dynamical equations on the gravity side. Specifically, transport coefficients of the dual medium, which have to be compared with their experimentally measured values, should follow, after applying the AdS/CFT computational machinery, from characteristics of the corresponding AdS Black Hole (BH) solutions \cite{Hartnoll:2008vx},\cite{Kovtun:2004de}. Other properties of the dual media such as the temperature and the charges density are also determined by characteristics of BHs in the AdS bulk.

If the correspondence between temperature and charge density on both sides of the Duality is clear (in thermodynamically equilibrium state and for stationary BH solutions temperature and charge density of a dual fluid are that of a BH), one may wonder how a BH characteristics influence on transport coefficients of the dual medium momentum/charge flows. A remarkable idea on the BH horizon as a stretched membrane \cite{MPbook} explains this relation: the stretched horizon is an effective ``fluid'' which is not an ideal one. Within the Membrane Paradigm scattering on a BH waves cause different flows when interact with the effective fluid near the horizon. Physical properties of this medium are encoded in characteristics of the BH  response to the background fluctuations \cite{Kovtun:2003wp}. For instance, the quasi-normal modes of a BH (see a review \cite{Konoplya:2011qq} and Refs. therein) are directly related to the shear viscosity \cite{Starinets:2002br} and the viscous relaxation time in the effective fluid within the hydrodynamic limit of AdS/CFT correspondence \cite{Kovtun:2005ev},\cite{Mas:2007ng}. Conformal symmetry of the AdS bulk translates the physical properties of the stretched horizon membrane to the AdS boundary, acting as the RG flow from the weak to the strong coupling constant and closing the Gauge/Gravity correspondence in this way. Therefore, properties of the dual strongly coupled media are determined by specifics of the BH horizons.

But talking on properties of a fictitious surface of the BH horizon we have only to keep in mind its geometric properties and nothing more. A difference in geometry of the horizon surfaces shall provide the difference in characteristics of dual fluids. A notable example of this point is introducing anisotropy in the BH solution \cite{Mateos:2011ix},\cite{Erdmenger:2010xm},\cite{Rebhan:2011vd} that generates anisotropy of the dual fluid flows: values of transport coefficients are different in different flow directions. Introducing anisotropy is a valuable step towards description of real QGP. We propose \cite{Moskalets:2014hoa} another step in this direction introducing the spatial ``inhomogeneity'' on the horizon surface.  
Let us briefly discuss effects of this generalisation.

\section{Inhomogeneity on the Gravity side 
and the dual fluid response} 

In \cite{Moskalets:2014hoa} we found a novel solution of a charged static AdS$_4$ black hole
\[
ds^2=-f(r)dt^2+\frac{dr^2}{f(r)}+r^2 e^{\Phi(x,y)}(dx^2+dy^2),\quad f(r)=\frac{r^2}{l^2}-\frac{\omega_2 M}{r}+K+\sum_{i=e,m}\,\frac{k^2 Q^2_{i}}{2r^2}\,,
\]
\begin{equation}
A_t=\mu-\frac{Q_e}{r},\qquad A_x=-Q_m \int e^{\Phi(x,y)}\,dy
\label{RNsolem}
\end{equation}
with the metric potential $\Phi(x,y)$. In dependence on the Gauss curvature of the horizon surface $K=0,\pm 1$, the metric potential obeys the elliptic wave/Liouville equation
\begin{equation}
\frac{\partial^2 \Phi}{\partial x^2}+\frac{\partial^2 \Phi}{\partial y^2}+2Ke^{\Phi(x,y)}=0\,.
\label{LeqK}
\end{equation}
$Q_{e,m}$ are the electric/magnetic charge densities; we refer the reader to \cite{Moskalets:2014hoa} for definitions of other quantities in (\ref{RNsolem}). The conformal factor $\Phi(x,y)$ carries the information on local internal geometry of the event horizon surface. We should note the fictive character of this ``surface'' and of the spatial inhomogeneity distribution on it as well. Nevertheless, thinking of the horizon as a stretched membrane helps to visualise these items. Since the metric potential
$\Phi(x,y)$ serves to differentiate local regions of the horizon from each other, it plays the role of the BH hair. The Liouville equation (\ref{LeqK}) constrains possible configurations of inhomogeneity distribution function, leaving however enough freedom to fit the gravity system configuration to the desired properties of the dual gauge system (see below).

It is straightforward to construct the higher-dimensional generalisation of the solution (\ref{RNsolem}) with the Liouville equation constraint  (\ref{LeqK}) (see \cite{Moskalets:2014hoa} for details); other higher-dimensional solutions, generalising (\ref{RNsolem}) to the metric potentials of more than two of the horizon coordinates, may be found in \cite{Hao:2014xva},\cite{Hao:2015zxa}. Note however, constraints on the metric potentials \cite{Hao:2014xva},\cite{Hao:2015zxa} do not generally admit exact solutions.

Computations of the transport coefficients in the hydrodynamic limit of AdS/CFT correspondence may be proceeded almost along the same line as before (see \cite{Policastro:2002se},\cite{Kovtun:2003wp},\cite{Kovtun:2004de}): the conformal flatness of the horizon justifies employing this technique. To avoid complications with BH phase transitions and/or thermodynamical instability we choose the solution (\ref{RNsolem}) with a planar inhomogeneous horizon ($K=0$). The charge diffusion coefficient follows from the diffusion pole of the retarded Green's function \cite{Policastro:2002se},\cite{Kovtun:2003wp}: it comes from a vector field fluctuations over the BH background along the horizon coordinates. But we should take into account there is not a symmetry of the background in $x,y$ directions. So we have to consider a  general (cf. \cite{Policastro:2002se},\cite{Kovtun:2003wp}) perturbation 
\begin{equation}
A_m(t,r,x,y)=\int \, \frac{d\omega d^2q}{(2\pi)^3} \, e^{-i\omega t+iq_{(x)}x+iq_{(y)}y}\,A_m(\omega,q_{(x)},q_{(y)},r)\,
\label{AansG}
\end{equation}
of the Maxwell field over the neutral (Schwarzschild-type) or over the charged (Raissner-N\"ordstrom-type) AdS$_4$ BH background. Solution to the Maxwell equations with the appropriate boundary conditions
is used to construct the retarded Green's function of small fluctuations (\ref{AansG}) on the boundary of AdS space. In the hydrodynamic limit of Gauge/Gravity Duality the diffusion pole of the vector mode retarded Green's function \cite{Policastro:2002se},\cite{Kovtun:2003wp},\cite{Kovtun:2004de} leads to the following value of the diffusion coefficient \cite{Moskalets:2014hoa} 
\begin{equation}
D=\frac{3}{4\pi T_{N,RN}}\, e^{-\Phi(x,y)}\,.
\label{DApert}
\end{equation}
$T_{N,RN}$ are the Hawking temperatures of the neutral  ($N$) or the charged ($RN$) AdS$_4$ BHs.

Computing the shear viscosity from the pole of the retarded Green's function of gravitational field fluctuations \cite{Policastro:2002se},\cite{Kovtun:2003wp} we have to choose the perturbation ansatz on account of asymmetry in $x,y$ directions. The consistency of the perturbed Einstein equation in the leading order approximation is achieved with choosing the perturbation as a sum of two modes, $\delta g_{mn}=h_{mn}+H_{mn}$, with $h_{mn}=\{ h_{ty}(t,r,x),h_{xy}(t,r,x)\}$ and $H_{mn}=\{ H_{tx}(t,r,y),H_{xy}(t,r,y)\}$.
Solving for the Einstein equation with in-coming boundary condition we find, after the standard machinery of AdS/CFT \cite{Aharony:1999ti}, the pole of the retarded Green's function corresponding to the momentum transfer in $x,y$ directions and hence to the shear viscosity-per-entropy density ratio:
\begin{equation}
D=\frac1{4\pi T}\,e^{-\Phi}\,,\quad \frac{\eta}{s_0}=DT=\frac1{4\pi}e^{-\Phi}\,.
\label{shear}
\end{equation}
As in the case of the charge diffusion (\ref{DApert}) the exponential suppression results in
violation of the corresponding universal bound (the KSS \cite{Kovtun:2004de} bound $\eta/s_0 \ge 1/4\pi$ in the case) in domains of positivity of the metric potential $\Phi(x,y)$. Note that experimentally measured values of $\eta/s$ in real Quark-Gluon Plasma \cite{Gale:2012rq},\cite{Song:2012ua} do not violate the KSS bound. Below we will turn to the discussion of this important point in more detail.

Higher-dimensional extension of our solutions (\ref{RNsolem}) produces the anisotropy of $d\ge 4$ effective dual media. One gets two types of diffusion coefficients in that case: the normal (cf., e.g., \cite{Policastro:2002se},\cite{Kovtun:2003wp}) and the suppressed (cf. \cite{Moskalets:2014hoa}) one. Last but not least, the flexibility in constructing exact solutions to the Liouville equations allows us to find a trial distribution function $\Phi(x,y)$ (see \cite{Moskalets:2014hoa} for details). Plugging it in (\ref{shear}), one recovers the range of theoretically predicted and experimentally measured values of $\eta/s$, from the KSS bound value $0.08$ to the highest value $0.2$ \cite{Song:2012ua}.


\vspace{-0.4cm}
\section{Summary}

We proposed novel solutions of static charged AdS BHs with inhomogeneity distribution on the horizon surface. The inhomogeneity is encoded in the metric potential, which depends on horizon coordinates and is constrained to obey the Liouville equation \cite{Moskalets:2014hoa}. It is important to note the metric potential plays the role of a BH hair. This observation helps to look at the obtained solutions and they role in the Gauge/Gravity Duality from a different angle.

Recall, a BH state with homogeneous horizon (a ``no hair'' state) is the very state with maximum of entropy. An elementary explanation of this fact may be found in L. Susskind's remarkable book \cite{Sbook}. Therefore, the standard RN black hole with trivial metric potential $\Phi(x,y)$ possesses more entropy than that of a charged BH with a non-trivial hair. We get two significant corollaries of this fact. First, physically accepted domains of the metric potential $\Phi(x,y)$ are those of its non-positive values. It comes from $s=s_0 \exp{\Phi(x,y)}$ (see \cite{Moskalets:2014hoa}) between entropy $s$ of the hairy BH and the entropy of the standard RN solution $s_0$.  Second, evolution of a physical system proceeds towards the entropy increasing. Therefore, within the Gauge/Gravity Duality a QGP may only evolve in the direction of $\eta/s_0$ decreasing, reaching the KSS bound value in the homogeneous state. Perhaps, this state is unreachable due to forming hadrons back faster than formation of a homogeneous medium. At least, we can explain the noticeable difference between the KSS bound value and the experimental data in this way. Since it would be naively to expect homogeneity of a QGP right after the collision event, we believe our results open a new window in studies of strongly coupled fluid of quarks and gluons.

\bigskip
\centerline{\bf Acknowledgements } 
TM is grateful to Organizers and participants of 43rd ITEP Winter School for kind hospitality and productive atmosphere during the School. AN appreciates Prof. M.I. Gorenstein for a valuable discussion on current status of QGP.

\end{document}